\documentclass[twocolumn,pra,showpacs,aps]{revtex4}

\usepackage{amsmath}
\usepackage{amsbsy}
\usepackage{amsthm}
\usepackage{amssymb}
\usepackage{latexsym}
\usepackage[dvips]{color,graphicx}

\begin{document}

\title{Hard-core bosons on optical superlattices:\\
Dynamics and relaxation in the superfluid and insulating regimes}

\author{Marcos Rigol}
\affiliation{Physics Department, University of California,
Davis, California 95616, USA}
\author{Alejandro Muramatsu}
\affiliation{Institut f\"ur Theoretische Physik III, Universit\"at Stuttgart, 
Pfaffenwaldring 57, D-70550 Stuttgart, Germany}
\author{Maxim Olshanii}
\affiliation{Department of Physics \& Astronomy,
University of Southern California, Los Angeles, California 90089, USA}

\begin{abstract}
We study the ground-state properties and nonequilibrium dynamics 
of hard-core bosons confined in one-dimensional lattices in 
the presence of an additional periodic potential (superlattice) 
and a harmonic trap. The dynamics is analyzed after a sudden 
switch-on or switch-off of the superlattice potential, 
which can bring the system into insulating or superfluid phases,
respectively. A collapse and revival of the zero-momentum peak can be 
seen in the first case. We study in detail the relaxation 
of these integrable systems towards equilibrium. We show how after 
relaxation time averages of physical observables, like the momentum 
distribution function, can be predicted by means of a generalization 
of the Gibbs distribution.
\end{abstract}
\maketitle

\section{Introduction}

In recent years the nonequilibrium dynamics of low-dimensional 
quantum systems has been increasingly attracting the attention 
of experimentalists and theoreticians from different areas of 
physics \cite{dresden06}. An exciting field in which great 
experimental progress has been achieved within the last decade 
is the one of ultracold quantum gases. There, advances in atom 
waveguide technology \cite{thyw99,muller99,dekker00,key00,bongs01}, 
the realization of quantum gases in very anisotropic traps 
\cite{schreck01,gorlitz01,strecker02}, and loading Bose-Einstein 
condensates (BEC's) on optical lattices 
\cite{greiner01,moritz03,stoferle03,tolra04,paredes04,kinoshita04,fertig05,kinoshita06}
have allowed experimentalists to obtain a wide variety of systems
in which the effects of the reduced dimensionality can be studied in 
detail. 

Due to the very high control than can be achieved in an experiment
with a degenerate quantum gas, one can prepare it under very specific 
initial conditions and study its evolution. This has been done in the 
one-dimensional (1D) regime in various experiments. For example, the 
transport properties of 1D Bose gases on a lattice have been studied in 
Refs.~\cite{stoferle03,fertig05}, where the gas was displaced from 
the center of the trap and allowed to oscillate. More recently, 
Kinoshita {\it et~al.} \cite{kinoshita06} have addressed experimentally 
the question of whether an isolated integrable or nearly integrable 
system can relax to the thermal equilibrium state \cite{rigol06_1}.

On the theoretical side, integrability in low-dimensional systems 
allows one to perform exact studies of the equilibrium properties 
and the nonequilibrium dynamics of well-known models 
\cite{igloi00,sengupta04,calabrese06,cherng06,cazalilla06}, 
some of which have already become relevant 
to experiments. One particular model in which we are interested in 
this work is the one of impenetrable bosons in 1D 
\cite{paredes04,kinoshita04,kinoshita06}. It has been shown 
theoretically \cite{olshanii98,petrov00,dunjko01} that 
in certain 1D regimes of low densities and low temperatures, 
bosons behave as a gas of impenetrable particles also known as 
Tonks-Girardeau bosons [hard-core bosons (HCB's)].

The 1D homogeneous gas of HCB's was introduced by Girardeau 
\cite{girardeau60}, who also established a one-to-one 
correspondence (Bose-Fermi mapping) between 1D HCB's and 
spinless fermions. This mapping has been recently used to 
study the nonequilibrium case 
\cite{girardeau00,girardeau00a,girardeau02,minguzzi05,campo05}, 
where the density dynamics revealed dark-soliton 
structures \cite{girardeau00}, breakdown of the time-dependent 
mean-field theory \cite{girardeau00a}, interference 
patterns of the thermal gas on a ring \cite{girardeau02},
and other interesting effects during the expansion 
\cite{minguzzi05,campo05}.

Hard-core bosons have been also realized experimentally in the presence 
of a lattice along the 1D tubes \cite{paredes04}. In the periodic case, 
the HCB lattice Hamiltonian can be mapped onto the 1D $XY$ model of Lieb, 
Schulz, and Mattis \cite{lieb61}. The 1D $XY$ model has been extensively 
studied in the literature, and the asymptotic behavior of the 
correlation functions is known \cite{mccoy68,vaidya78,mccoy83,gangardt06}.
With an additional confining potential, the case relevant to the 
experiments in Ref.~\cite{paredes04}, this model has been studied 
be means of an exact numerical approach in Ref.~\cite{rigol04_1}. 
There it was shown that one-particle correlations exhibit a universal 
power-law decay. The generalization of the approach in Ref.~\cite{rigol04_1} 
to the nonequilibrium dynamics \cite{rigol04_2,rigol05_1,rigol05_2} 
revealed that during the expansion
of the HCB gas on the lattice two very different regimes can be identified.
If the expansion starts from a Mott insulating state, quasi-long range 
correlations develop between initially uncorrelated particles, producing
the emergence of quasicondensates at finite momentum 
\cite{rigol04_2,rigol05_2}. On the other hand, for low initial 
densities (superfluid state), the momentum distribution of expanding 
HCB's rapidly approaches that of noninteracting fermions 
\cite{rigol05_1,rigol05_2}.

In this work we study the nonequilibrium dynamics of hard-core bosons
on 1D superlattices. The superlattice is obtained adding an 
extra periodic potential to the already existing lattice. In the 
soft-core regime these systems have been already realized experimentally 
\cite{peil03,strabley06} and studied theoretically in 1D by various 
mean-field and perturbative approaches \cite{buonsante04}, 
quantum Monte Carlo simulations \cite{rousseau06}, 
and exact diagonalization \cite{rey06}. In the hard-core regime, 
due to the mapping to noninteracting fermions, 
one can realize that the effect of the extra periodic potential 
is to open gaps at the boundaries of the reduced Brillouin zone. 
This means that in addition to the insulating phase with density 
1 (full filling) new insulating phases appear at fractional fillings. 

In three-dimensional systems the study of the half-filled hard-core 
boson model allowed one to prove rigorously the existence of 
Bose-Einstein condensation and Mott insulating phases
tuning the strength of the additional lattice \cite{aizenman04}. 
Here we quench the strength of the superlattice potential to study the 
dynamics of these systems in the superfluid and insulating regimes.
We are interested in understanding how the system approaches 
equilibrium, if it does, and in testing the prediction power of a 
generalized Gibbs ensemble theory recently proposed in Ref.\ \cite{rigol06_1}.
In order to do so, we also study the case in which an additional harmonic 
confining potential is introduced in the system, as relevant to most of the
experimental situations. At low densities we obtain results similar 
to that of the recent experiment by Kinoshita {\it et al.} \cite{kinoshita06}. 

The exposition is organized as follows. In Sec.\ II we describe the 
ground-state properties of HCB's in a periodic superlattice, paying 
special attention to the behavior of the one-particle correlations in 
the superfluid and insulating regimes. Section III is devoted to the study
of the nonequilibrium dynamics of the system after a sudden switch-on 
and -off of the superlattice potential---i.e., when going from the superfluid 
to the insulating phases and vice versa. Relaxation and the 
collapse and revival of the zero-momentum peak of the momentum 
distribution function are studied in detail.
In Sec.\ IV we analyze the ground-state properties of the system in the
presence of a trap, which produces a coexistence of superfluid and 
insulating phases. The nonequilibrium dynamics in the presence of 
a trap is studied in Sec.\ V. Finally, the conclusions are presented
in Sec.\ VI.

\section{Hard-core bosons in periodic superlattices}

The hard-core boson Hamiltonian in the presence of a superlattice 
potential can be written as
\begin{equation}
\label{HamHCB} \hat{H} = -\sum_{i} \left( t_{i,i+1} \hat{b}^\dagger_{i} 
\hat{b}^{}_{i+1} + \mathrm{H.c.} \right) 
+ A \sum_{i} \cos \left( {2 \pi i \over L} \right) \hat{n}_i,
\end{equation}
where the HCB creation and annihilation operators at site $i$ 
are denoted by $\hat{b}^{\dagger}_{i}$ and $\hat{b}^{}_{i}$, respectively, 
and the local density operator by $\hat{n}_i=\hat{b}^{\dagger}_{i}\hat{b}^{}_{i}$.
In different sites the HCB creation and annihilation operators
commute as usual for bosons:
\begin{equation}
[\hat{b}^{}_{i},\hat{b}^{\dagger}_{j}]=
[\hat{b}^{}_{i},\hat{b}^{}_{j}]=
[\hat{b}^{\dagger}_{i},\hat{b}^{\dagger}_{j}]=0,\quad \mathrm{for} \quad i\neq j.
\end{equation}
However, on the same site these operators satisfy anticommutation 
relations typical for fermions:
\begin{equation}  
\left\lbrace  \hat{b}^{}_{i},\hat{b}^{\dagger}_{i}\right\rbrace =1, 
\qquad
\hat{b}^{\dagger 2}_{i}= \hat{b}^2_{i}=0.
\label{ConstHCB} 
\end{equation}
These constraints avoid double or higher occupancy of the 
lattice sites. In Eq.\ (\ref{HamHCB}), the hopping parameters 
are denoted by $t_{i,i+1}=t$. The last term represents the 
superlattice potential with strength $A$ and unit cells with 
$L$ sites.

Using the Jordan-Wigner transformation \cite{jordan28}
\begin{equation}
\label{JordanWigner} \hat{b}^{\dag}_i=\hat{f}^{\dag}_i
\prod^{i-1}_{\beta=1}e^{-i\pi \hat{f}^{\dag}_{\beta}\hat{f}^{}_{\beta}},\qquad
\hat{b}_i=\prod^{i-1}_{\beta=1} e^{i\pi \hat{f}^{\dag}_{\beta}\hat{f}^{}_{\beta}}
\hat{f}_i \ ,
\end{equation}
one can map the HCB Hamiltonian onto the one of 
noninteracting spinless fermions,
\begin{eqnarray}
\label{HamFerm} \hat{H}_F =-\sum_{i} \left(t'_{i,i+1} \hat{f}^\dagger_{i}
\hat{f}^{}_{i+1} + \mathrm{H.c.} \right)+ A \sum_{i} \cos 
\left( {2 \pi i \over L} \right) \hat{n}^f_i,
\end{eqnarray}
where $\hat{f}^\dagger_{i}$ and $\hat{f}_{i}$ are the creation and
annihilation operators for spinless fermions at site 
$i$ and $\hat{n}^f_{i}=\hat{f}^\dagger_{i}\hat{f}^{}_{i}$ is the local 
particle number operator. Considering that
\begin{equation}
\hat{b}^\dagger_{N} \hat{b}^{}_{1}=-\hat{f}^\dagger_{N}\hat{f}^{}_{1}\ 
\exp\left(- i\pi \sum^N_{\beta=1}\hat{n}^f_{\beta}\right) ,
\end{equation}
where $N$ is the number of lattice sites, one notices that
$t'_{N,1}=\exp\left[-i\pi (N_b+1)\right] t_{N,1}$; i.e., 
for periodic chains when the number of particles in the system 
($N_b=\sum_{i}\langle \hat{n}_i \rangle=\sum_{i}\langle \hat{n}^f_i\rangle$)
is odd, the equivalent fermionic Hamiltonian satisfies the same 
boundary conditions than the HCB one; otherwise, if $N_b$ is even, 
the opposite boundary conditions are required for the fermionic
Hamiltonian.

With the help of the mapping above one can realize that for HCB's in 
a superlattice insulating phases appear at fractional fillings 
($n_i=i/L$, with $i=1,\ldots, L-1$, unless a band crossing occurs 
\cite{rousseau06}), in addition to the full filling 
insulator ($n_i=1$) already present in the absence of the superlattice. 
This is because in the equivalent system of noninteracting fermions 
the superlattice potential opens gaps at the boundaries of the reduced 
Brillouin zones. In what follows, for simplicity, we restrict our study 
to superlattices with $L=2$. In this case the insulating phases occur at
half and full filling. The generalization of our results to larger values
of $L$ is straightforward, and does not (qualitatively) change the 
results we discuss for $L=2$.

In order to study the bosonic one-particle correlations
\begin{equation}
\rho_{ij}=\left\langle \hat{b}^\dagger_{i}\hat{b}_{j}\right\rangle 
\end{equation} 
and related quantities like the momentum distribution function
 \begin{equation}
n_k=\dfrac{1}{N}\sum_{jl} e^{-ik(j-l)}\rho_{jl},
\end{equation}
we follow the exact approach described in detail in 
Ref.\ \cite{rigol04_1}. This approach allows us to study very large 
system sizes (thousands of lattice sites) in a very efficient way.

\begin{figure}[h]
\begin{center}
\includegraphics[width=0.44\textwidth]
{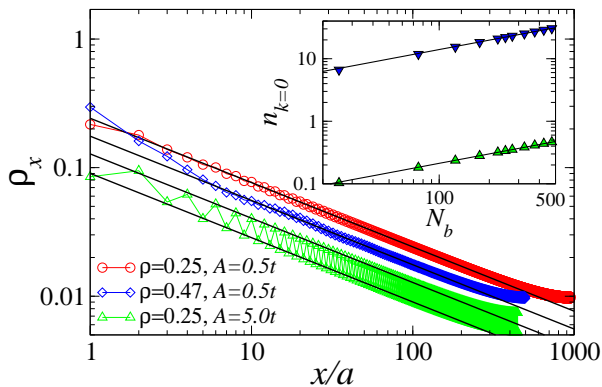}
\end{center} \vspace{-0.7cm}
\caption{(Color online) Decay of one-particle correlations 
(averaged per unit cell as described in the text) in the superfluid phase. 
These systems have up to 1900 lattice sites and $L=2$. The plot for 
$\rho=0.25$, $A=5t$ was displaced down for clarity. Straight solid 
lines correspond to power laws $1/\sqrt{x}$. The inset shows $n_{k=0}$ (top)
and $n_{k=\pm\pi/a}$ (bottom) vs $N_b$ for systems at quarter filling and $A=0.5t$. 
The straight lines signal $\sqrt{N_b}$ behavior. Distances are normalized 
by the lattice constant $a$.}
\label{DensityMatrix_Superfluid}
\end{figure}

In Fig.\ \ref{DensityMatrix_Superfluid} we show how the 
one-particle correlations decay in the presence of a superlattice 
potential in the superfluid phase. In the figure we have averaged
the correlations measured from the even and odd sites in order 
to minimize the effects of the different density in the sites; 
i.e., we have plotted
\[
\rho_x=\frac{1}{2}\left( \rho_{i^{odd},i^{odd}+x/a}+
\rho_{i^{even},i^{even}+x/a}\right).
\]
As can be seen in Fig.\ \ref{DensityMatrix_Superfluid}, 
one-particle correlations in the superlattice decay with the same power 
law $\rho_x\sim 1/\sqrt{x}$ that was shown to be universal in absence 
of the superlattice potential \cite{rigol04_1}. The only effect that 
the superlattice introduces and that is evident only for large values 
of $A$ is an oscillatory behavior in $\rho_x$ on top of the $1/\sqrt{x}$
decay.

\begin{figure}[h]
\begin{center}
\includegraphics[width=0.48\textwidth]
{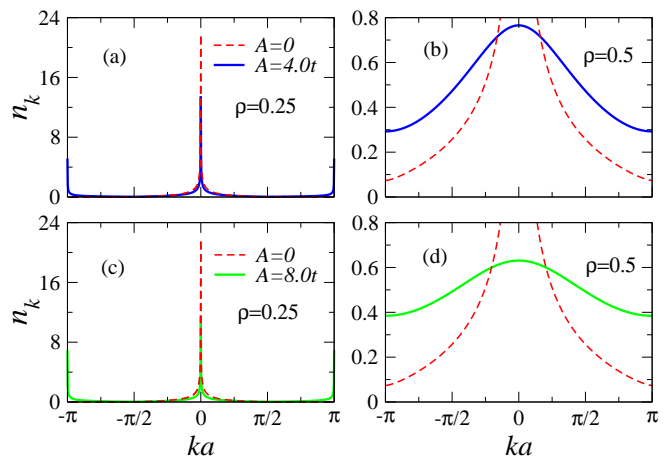}
\end{center} \vspace{-0.7cm}
\caption{(Color online) Momentum profiles for periodic systems with 900 lattice 
sites in a superlattice potential with $L=2$. In (a) and (c) the system is in a  
superfluid state. The peaks at $k=0$ and $k=\pm \pi/a$ signal the presence
of quasi-long range one-particle correlations (Fig.\ \ref{DensityMatrix_Superfluid}). 
On the other hand, in (b) and (d) the system is in an insulating state where 
one-particle correlations decay exponentially (Fig.\ \ref{DensityMatrix_Insulator}).}
\label{PerfilK_Periodic}
\end{figure}

The existence of quasi-long-range one-particle correlations implies that, 
like in the usual lattice, a sharp peak in the momentum distribution 
function at $k=0$ signals the superfluid state. The additional 
oscillatory behavior seen in $\rho_x$ (Fig.\ \ref{DensityMatrix_Superfluid}) 
is reflected by additional peaks in $n_k$ at $ka=\pm\pi$, 
which deplete the one in $k=0$ from its value for $A=0$. This can be seen 
in Figs.\ \ref{PerfilK_Periodic}(a) and \ref{PerfilK_Periodic}(c), 
where we have plotted the momentum distribution function for two different 
values of the strength of the superlattice potential. A simple calculation
also allows to extract from $\rho_x\sim 1/\sqrt{x}$ the scaling behavior
of the $k=0$ and $k=\pm\pi/a$ peaks as the system size is increased 
keeping the density constant. One finds that $n_{k=0,\pm \pi/a}\sim \sqrt{N_b}$. 
Such scaling can be seen in the inset of Fig.\ \ref{DensityMatrix_Superfluid} 
where we have plotted these quantities for systems at quarter filling 
and $A=0.5t$.

At half filling the behavior of the system is completely different to the 
one above. As mentioned before a gap opens in the spectrum. 
Since Eq.\ (\ref{HamFerm}) can be easily diagonalized for $L=2$, 
one immediately obtains that the gap is $\Delta=2A$, and the dispersion 
relation in the two bands is
\begin{equation}
\epsilon_{\pm}(k) = \pm \sqrt{ 4 t^2 \cos^2(ka) + A^2} ,
\end{equation}
where by ``$+$'' we mean the upper band and by ``$-$'' the lower one.

\begin{figure}[h]
\begin{center}
\includegraphics[width=0.44\textwidth]
{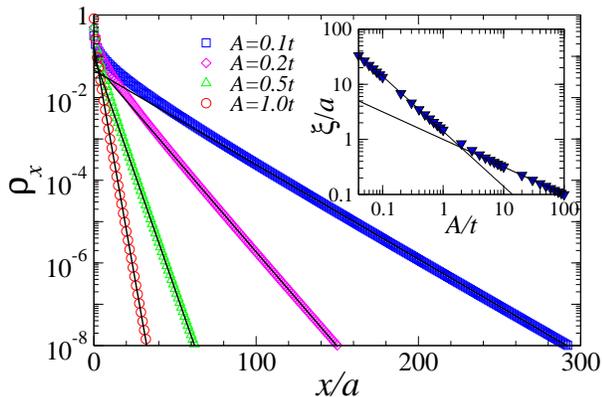}
\end{center} \vspace{-0.7cm}
\caption{(Color online) Decay of one-particle correlations 
(averaged per unit cell as described in the text) in the insulating
(half-filled) case. The systems considered have up to 1500 lattice 
sites and $L=2$. The straight lines signal an exponential 
decay with a correlation length $\xi$, which is a function 
of $A$. In the inset we show $\xi$ vs $A$. The straight 
lines depict the asymptotic behavior of $\xi$. For very small values 
of $A$ one has that $\xi/a\sim 1/(A/t)$ while for large values of $A$ 
one obtains that $\xi/a\sim 1/\sqrt{A/t}$ (where $a$ is the lattice 
constant).}
\label{DensityMatrix_Insulator}
\end{figure}

The presence of this gap produces an exponential decay 
of the one-particle correlations as shown in 
Fig.\ \ref{DensityMatrix_Insulator}. The 
correlation length $\xi$ is a function of the gap---i.e., 
of $A$. We have calculated the correlation length $\xi$ 
as the second moment of the one-particle density matrix,
\begin{equation}
\label{secmom}
\xi=\sqrt{\dfrac{1}{2}
\dfrac{\sum_{ij}\left( x_i-x_j \right)^2 \rho_{ij}}{\sum_{ij} \rho_{ij}}},
\end{equation}
which for large values of $\xi$ is equivalent to calculating it fitting an
exponential decay $\rho_x\sim \exp(-x/\xi)$, as shown at finite temperatures in
Ref.\ \cite{rigol05_3}.

The behavior of $\xi$ as a function of $A$ is shown in the inset of 
Fig.\ \ref{DensityMatrix_Insulator}. There one can see that it exhibits 
two different functional forms for small and large values of $A$. For
small values of $A$ we find that $\xi/a\sim 1/(A/t)$, while for large values 
of $A$ it is $\xi/a\sim 1/\sqrt{A/t}$ ($a$ is the lattice constant).
In general, a correlation length $\xi \sim 1/A$ develops from free 
bosonic excitations across a gap $\sim A$. Such an argument results 
already at a mean-field level, and in the absence of diverging correlation 
lengths, it is qualitatively correct for small values of $A$.
For very large values of $A$ one can use perturbation theory to determine 
the asymptotic behavior of $\xi$. For $A/t\rightarrow \infty$ the ground 
state at half filling is just an array of Fock states with one particle 
in the odd sites and no particle in the even sites 
$\vert \Psi \rangle_{A/t\rightarrow \infty} =
\prod_i \vert 1 \rangle_{i_{odd}}\vert 0 \rangle_{i_{even}}$. Up to 
first order in perturbation theory the ground state of the system can be 
written as
\begin{equation}
\vert \Psi \rangle_G\approx \vert \Psi \rangle_{A/t\rightarrow \infty} +
\frac{t}{2A} \sum_{i}\left(\hat{b}^{\dagger}_{i}\hat{b}_{i+1}+\mathrm{H.c.}\right) 
\vert \Psi \rangle_{A/t\rightarrow \infty},
\end{equation}
which means that for $i\neq j$ [the only ones that contribute to the 
numerator in Eq.\ (\ref{secmom})]
\begin{eqnarray}
\rho_{ij}=\langle \hat{b}^{\dagger}_{i}\hat{b}_{j} \rangle_G
&=& \frac{t}{2A} \quad \mathrm{for}\ \  |i-j|=1, \nonumber \\
&=& 0 \qquad \mathrm{for}\ \ |i-j|>1.
\end{eqnarray} 
In the denominator in Eq.\ (\ref{secmom}) only the terms $i=j$ (the densities)
contribute to first order. Hence, 
\begin{equation}
\xi=a\sqrt{\frac{t}{A}}
\end{equation}
for very large values of $A/t$. 
Indeed, $\xi/a=\sqrt{t/A}$ is the straight line 
that in the inset in Fig.\ \ref{DensityMatrix_Insulator} follows the 
data points for large values of $A$.

The consequence of the exponential decay of the one-particle correlations 
in the momentum distribution function can be seen in 
Figs.\ \ref{PerfilK_Periodic}(b) and \ref{PerfilK_Periodic}(d). There
we have plotted $n_k$ for two different values of $A$ and compared it 
with $A=0$. The effect of the gap at half filling is dramatic. It destroys
the peaks in the momentum distribution function. This feature can be used 
experimentally to detect the presence of an insulating state in a 
superlattice, like it has been done previously to detect the presence
of a Mott insulator in the absence of the superlattice. 

\begin{figure}[h]
\begin{center}
\includegraphics[width=0.44\textwidth]
{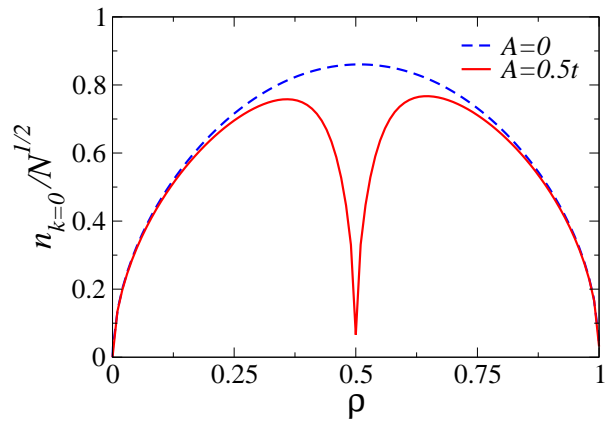}
\end{center} \vspace{-0.7cm}
\caption{(Color online) Normalized occupation of the $k=0$ 
momentum state as a function of the density in periodic systems 
with $N=1000$. For $A\neq 0$ one can see that $n_{k=0}$ is strongly 
suppressed around $n=0.5$, where the system is an insulator.}
\label{nk_0vsNb}
\end{figure}

An overall picture of the behavior of the momentum 
distribution function when changing the density can be obtained plotting
$n_{k=0}$ vs $\rho$ as shown in Fig.\ \ref{nk_0vsNb}. For contrast we have 
also plotted the result for $A=0$. Figure \ref{nk_0vsNb} shows that while 
for low densities the effect of $A$ is almost imperceptible, it becomes 
very large as the density approaches half filling. In the insulating 
state $n_{k=0}$ attains its minimum value.

\section{Dynamics and relaxation in the 
superfluid and insulating regimes}

We study in this section the dynamics of hard-core bosons
on superlattices when by a sudden change of the strength 
of the superlattice potential one goes from a superfluid to an 
insulating regime, and vice versa. Since for $L=2$ this can only 
occur at half filling, we will restrict our analysis to that case. 
In addition, for a closer connection with the experiments instead 
of periodic systems we consider open boxes, where translational 
invariance is broken, but keeping the same phase diagram than in the
periodic case. We will consider systems confined in harmonic traps, 
where superfluid and insulating phases can coexist, 
in the next two sections.

We start our study with the case in which the initial state is 
superfluid, at half filling with $A=0$, and $A$ is suddenly changed
to a finite value, for which in the ground state the system would be 
an insulator. This is similar, in the Hubbard model, to a change of 
the on-site repulsion $U$ from a value in which the system is 
superfluid to $U>U_c$ ($U_c$ being the critical value for the 
formation of a Mott insulator). The number of particles per site 
in this case has to be integer. Such study for a Hubbard like 
experimental system has been done in three-dimensional optical lattices 
in Ref.\ \cite{greiner02}.

In the experiment \cite{greiner02} it was found that as the system
evolves a collapse and revival of the initial momentum distribution 
function (the interference pattern) occurs. This can be easily 
understood considering that for time scales much smaller than the 
one set by the hopping parameter, the particle number per lattice 
site remains almost unchanged and only the phases evolve. 
This evolution is dictated by the on-site interaction 
between atoms ($U$), which in the Hubbard language reads
\begin{equation}
\hat{H}_{int}=\frac{1}{2} U \hat{n}(\hat{n}-1).
\end{equation}
Hence, the evolution operator has the form $\exp[-i U n (n-1)\tau/2]$, where
$\tau$ is a time variable (which we will give in units of $\hbar$ in what 
follows). This means that phases in different lattices sites evolve differently, 
according to the number of atoms present, collapsing the interference pattern. 
However, for times $\tau_{rev}=2\pi n/U$, where $n$ is an integer, a revival of the 
interference pattern occurs.

\begin{figure}[h]
\begin{center}
\includegraphics[width=0.44\textwidth]
{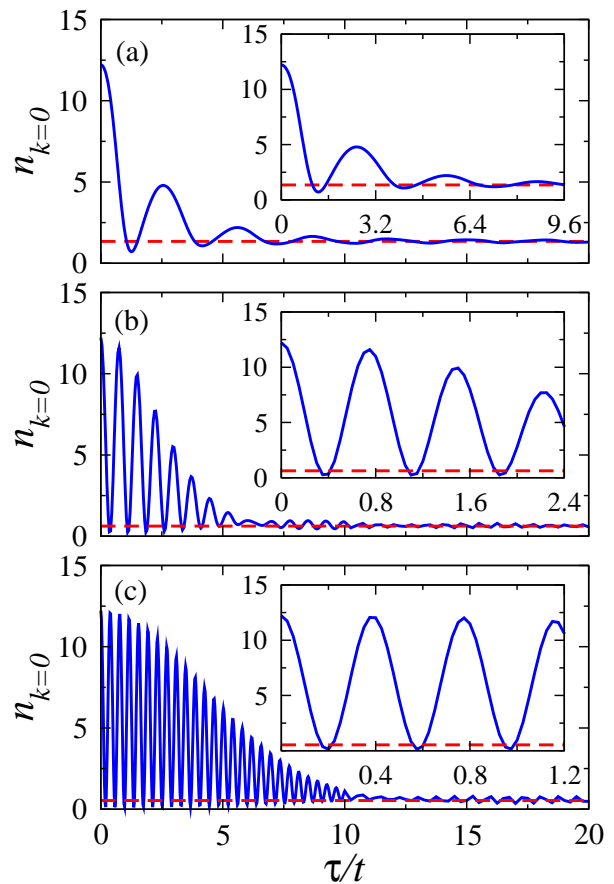}
\end{center} \vspace{-0.7cm}
\caption{(Color online) Evolution of the occupation of the zero-momentum 
state after $A$ is changed from $A=0$ in the initial (superfluid) state 
to $A=1.0t$ (a), $A=4.0t$ (b), and $A=8.0t$ (c). The insets show in more 
detail the first three revivals of $n_{k=0}$. Dashed lines show $n_{k=0}$ 
obtained from the constrained thermodynamics theory (see text).
These systems have 300 lattice sites, 150 particles, and $L=2$.}
\label{LatticeOn_nk0vstau}
\end{figure}

In Fig.\ \ref{LatticeOn_nk0vstau} we show the time evolution of the 
occupation of the zero-momentum state for a box that initially has 
$A=0$ and the evolution is performed by a sudden change to three 
different values of the final $A$. This figure shows that also in 
a 1D superlattice loaded with HCB's a collapse and revival of the 
zero-momentum peak occurs. Here, double or higher occupancy is 
forbidden by the hard-core constraint ($U=\infty$) so that the 
short-time evolution is only determined by the term in the 
Hamiltonian proportional to the superlattice potential,
\begin{equation}
\hat{H} = A \sum_{i} (-1)^i \hat{n}_i, 
\label{atomiclimit}
\end{equation}
which means that $A$ is the parameter that controls the 
revival time (equivalent to $U$ in the Hubbard case). This can be better 
seen in the insets in Fig.\ \ref{LatticeOn_nk0vstau}, where we have plotted 
the first three revivals of $n_{k=0}$. Comparing the times with the values
of $A$ one can see that indeed $\tau_{rev}\sim\pi n/A$.

The second effect that is apparent in Fig.\ \ref{LatticeOn_nk0vstau} 
is the damping of the collapse and revival of the zero-momentum peak 
of $n_k$. It is related to the change of the occupation of each lattice 
site due to the finite value of the hopping parameter $t$. Variations in 
the density destroy the periodic evolution set
by Eq.\ (\ref{atomiclimit}). In these systems the dynamics of the density 
is determined by a combination of the times scales given by $t$ and $A$. 
(Notice that $A$ in the even sites acts like a barrier between odd sites.)
In Fig.\ \ref{LatticeOn_nk0vstau} one can see that the largest damping 
occurs for the smallest the value of $A/t$. For the largest value of
$A$ shown in the figure ($A=8t$) the collapse and revival is almost completely
damped after $\tau=10t$.

After observing the damping of the collapse and revival of the momentum
distribution function in Fig.\ \ref{LatticeOn_nk0vstau} one immediate 
question one could ask is to what kind of momentum distribution is the 
system relaxing. In nonintegrable systems one expects $n_k$ to 
relax to the thermal distribution. In the grand canonical ensemble 
it can be obtained using the many-body density matrix 
\begin{equation}
\hat{\rho}=Z^{-1}\exp{\left[ -\left( \hat{H}-\mu \hat{N}_b\right) /k_BT\right]},
\end{equation}
where
\begin{equation}
Z=\mathrm{Tr}\left\lbrace 
\exp{\left[ -\left( \hat{H}-\mu \hat{N}_b\right)/k_BT\right]}\right\rbrace
\end{equation}
is the partition function. In order to determine the relevant 
temperature ($k_BT$) and the chemical potential ($\mu$) 
one can use the energy and number of particles of the evolving system, 
which do not change during the dynamics; i.e., $k_BT$ and $\mu$ 
can be calculated using the constraints
\begin{equation}
E=\mathrm{Tr}\left[ \hat{H}\hat{\rho} \right], \quad
N_b=\mathrm{Tr}\left[ \hat{N}_b\hat{\rho} \right].
\end{equation}

\begin{figure}[h]
\begin{center}
\includegraphics[width=0.44\textwidth]
{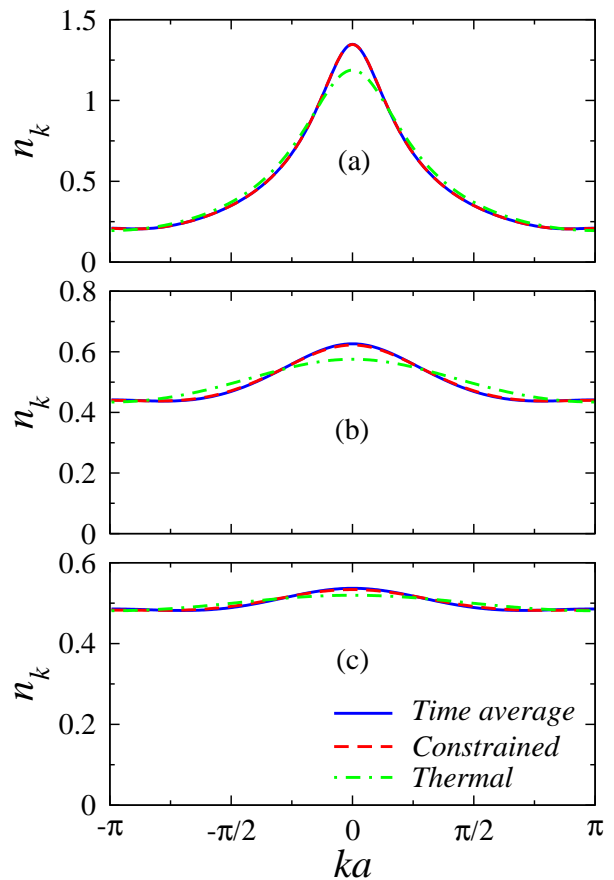}
\end{center} \vspace{-0.7cm}
\caption{(Color online) Time average of the momentum distribution function after the
damping of the oscillations seen in Fig.\ \ref{LatticeOn_nk0vstau}. 
We averaged measurements done in time intervals $\Delta\tau=20t$ 
between times $\tau=20t$ and $\tau=5000t$. The time average is 
compared with the results obtained in the usual thermal ensemble 
and the constrained theory explained in the text. These systems 
have 300 lattice sites, 150 particles, and $L=2$. In all cases
the initial $A=0$ and the final one $A=1.0t$ (a), $A=4.0t$ (b), and
$A=8.0t$ (c). The corresponding temperatures in the grand-canonical
ensemble are $k_BT=0.86t$ (a), $k_BT=6.86t$ (b), and $k_BT=25.75t$ (c).}
\label{LatticeOn_PerfilK}
\end{figure}

In Fig.\ \ref{LatticeOn_PerfilK} we compare the momentum distribution function
obtained within the grand-canonical approach described above (called ``thermal'' 
in the figure) with the time-averaged momentum distribution obtained during 
the evolution, after the damping of the oscillations shown in 
Fig.\ \ref{LatticeOn_nk0vstau}. We have averaged the time evolving 
$n_k(\tau)$ because, as can be seen in Fig.\ \ref{LatticeOn_nk0vstau}, 
this quantity exhibits fluctuations around its mean value. The exact $n_k$'s for
the thermal equilibrium were obtained using the method detailed in 
Ref.\ \cite{rigol05_3}. Figure \ref{LatticeOn_PerfilK} clearly shows that 
the mean values around which $n_k(\tau)$ fluctuates are not the ones 
of the usual thermal equilibrium; i.e., these systems are not ergodic 
in the usual sense. Since hard-core bosons in 1D lattices are integrable 
\cite{lieb61}, nonergodic behavior may have been expected due to 
the presence of extra constants of the motion (whose number is infinite 
in the thermodynamic limit) in addition to the energy and the number of 
particles.

The problem of relaxation in an integrable system, like the one of hard-core 
bosons that we are analyzing here, has been already discussed in 
Ref.\ \cite{rigol06_1}. There it was conjectured that the correct many-body 
density matrix that describes the properties of an integrable 
system after relaxation is given by a generalization of the 
Gibbs ensemble
\begin{equation}
\hat{\rho}_c=Z_c^{-1}\exp{\left( -\sum_m \lambda_m \hat{I}_m\right)},
\label{constrained}
\end{equation}
where
\begin{equation}
Z_c=\mathrm{Tr}\left[ 
\exp{\left( -\sum_m \lambda_m \hat{I}_m\right)}\right]
\label{constraints}
\end{equation}
is the corresponding partition function, $\lbrace I_m \rbrace$ is
a full set of integrals of motion, $\lbrace \lambda_m \rbrace$ are 
the Lagrange multipliers, and ${m}=1,\ldots,N$. 
These Lagrange multipliers can be calculated 
using the expectation values of the full set of integrals of motion 
of the evolving system, which do not change with time
\begin{equation}
\langle \hat{I}_m\rangle_\tau =
\mathrm{Tr}\left[ \hat{I}_m\hat{\rho}_c \right].
\label{multipliers}
\end{equation}
In Eq.\ (\ref{multipliers}) and in what follows, $\langle \cdots\rangle_\tau$
means expectation values in the time-evolving system when they do not depend
on time.

The conjecture above is still based on the ergodic hypothesis, but generalized 
to consider that the region available of phase space is determined by {\it all}
constants of the motion, and not only by $E$ and $N$ as usual for nonintegrable 
systems. Hence, what the density matrix defined by Eq.\ (\ref{constrained}) 
does is to maximize the many body Gibbs entropy,
\begin{equation}
S=k_B \mathrm{Tr}\left[ \hat{\rho}_c 
\ln\left( \frac{1}{\hat{\rho}_c}\right) \right],
\end{equation}
subject to the constraints imposed by all the integrals of motion.

Like in the ground state \cite{rigol04_1} and the thermal equilibrium 
case \cite{rigol05_3}, in order to calculate the HCB expectation values 
using Eq.\ (\ref{constrained}) we take advantage of the Jordan-Wigner 
transformation and the mapping to noninteracting fermions. In 
fermionic language the integrals of motion can be easily constructed 
after diagonalizing the Hamiltonian (\ref{HamFerm}):
\begin{equation}
\hat{H}_F \hat{\gamma}^{f\dagger}_m \vert 0\rangle= 
E_m \hat{\gamma}^{f\dagger}_m \vert 0\rangle.
\end{equation}
Since the occupation of the eigenstates of the final fermionic Hamiltonian
cannot change in time, these fermions are noninteracting; a complete set
of integrals of motion can be constructed to be
\begin{equation}
\left\lbrace \hat{I}^f_m\right\rbrace =
\left\lbrace \hat{\gamma}^{f\dagger}_m \hat{\gamma}^{f}_m\right\rbrace .
\label{constraintsf}
\end{equation}

The constraints in Eq.\ (\ref{constraints}), in the fermionic representation, 
lead to an analytical expression for the Lagrange multipliers,
\begin{equation}
\lambda_m=\ln\left( \frac{1-\langle\hat{I}^f_m\rangle_\tau}
{\langle\hat{I}^f_m\rangle_\tau}\right),
\end{equation}
which allows us to build the equivalent fermionic Hamiltonian and to
obtain the HCB expectation values using the approach explained in 
detail in Ref.\ \cite{rigol05_3}. One should notice at this point 
that the constraints defined by Eq.\ (\ref{constraintsf}), when  
written in the bosonic language, involve many bosonic creation 
and annihilation operators. Hence, these constraints lose the 
bilinear character they have in the fermionic representation.

\begin{figure}[b]
\begin{center}
\includegraphics[width=0.44\textwidth]
{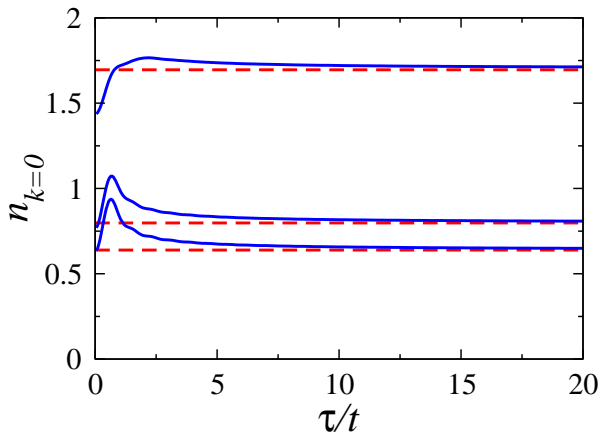}
\end{center} \vspace{-0.7cm}
\caption{(Color online) Evolution of the occupation of the zero-momentum 
state after $A$ is changed from $A=1.0t$, $A=4.0t$, and $A=8.0t$, 
(from top to bottom) in the initial (insulating) state to $A=0$. 
Dashed lines show $n_{k=0}$ obtained from the constrained thermodynamics 
theory (see text). These systems have 300 lattice sites, 150 particles, 
and $L=2$.}
\label{LatticeOff_nk0vstau}
\end{figure}

In Fig.\ \ref{LatticeOn_PerfilK} one can see that the results obtained
with the constrained thermodynamics theory explained above are 
indistinguishable from the ones of the time average after damping.
Hence, even for the evolution of these pure quantum states in integrable 
systems one can define a constrained ensemble able to predict the mean 
values of observables after relaxation. We have also included in 
Fig.\ \ref{LatticeOn_nk0vstau} the value of $n_{k=0}$ obtained from the
constrained thermodynamic theory for a comparison with the time evolving 
$n_{k=0}(\tau)$. There one can see that after relaxation the small 
fluctuations of $n_{k=0}(\tau)$ indeed occur around the value obtained 
with the generalized Gibbs ensemble. Notice that the case described 
so far in this section seems to be one of the worst scenarios for HCB's
since the evolution is performed with a Hamiltonian for which the ground
state of the system is gapped; i.e., relaxation is expected to 
occur slowly due to the presence of the gap.

In what follows we consider the inverse case. The case in which initially 
the system is in an insulating state and by turning off the superlattice 
potential this state is allowed to evolve with a Hamiltonian whose 
ground state is superfluid.

\begin{figure}[b]
\begin{center}
\includegraphics[width=0.432\textwidth]
{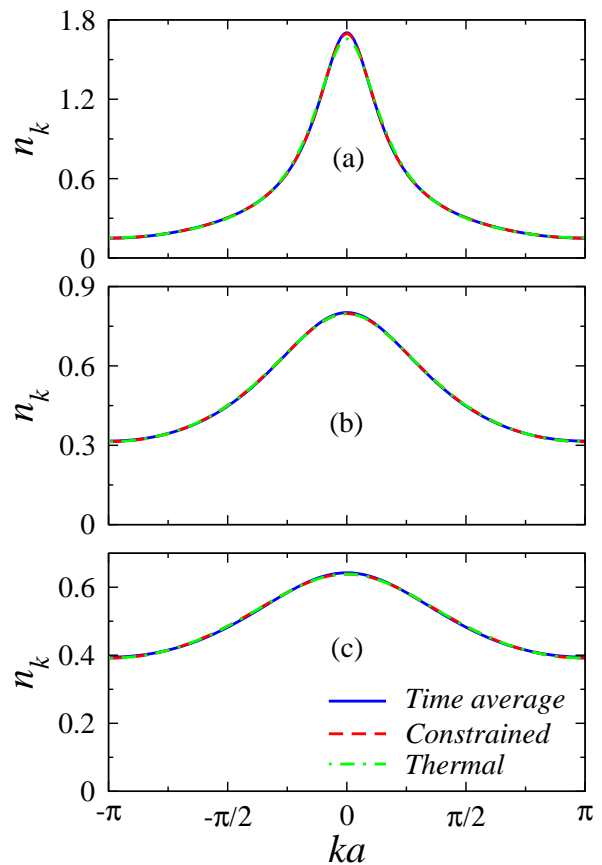}
\end{center} \vspace{-0.7cm}
\caption{(Color online) Time average of the momentum distribution 
function measured in time intervals $\Delta\tau=20t$ between times 
$\tau=20t$ and $\tau=5000t$. The time average is compared with the 
results obtained in the usual thermal ensemble and the constrained 
theory explained in the text. These systems have 300 lattice sites, 
150 particles, and $L=2$. Initially $A=1.0t$ (a), $A=4.0t$ (b), and
$A=8.0t$ (c), and in all cases the final value of $A$ is $A=0$. 
The corresponding temperatures in the grand-canonical ensemble are 
$k_BT=0.63t$ (a), $k_BT=2.06t$ (b), and $k_BT=4.03t$ (c).}
\label{LatticeOff_PerfilK}
\end{figure}

In Fig.\ \ref{LatticeOff_nk0vstau}, we show the time evolution of 
$n_{k=0}$ from three initial insulating states. They have the same 
values of $A$ used during the evolution of the systems in 
Figs.\ \ref{LatticeOn_nk0vstau} and \ref{LatticeOn_PerfilK}. The 
evolution is performed after turning off the superlattice potential 
(making $A=0$). In contrast to the case in which the superlattice 
is turned on, one can see in Fig.\ \ref{LatticeOff_nk0vstau} that 
no collapse and revivals occur in $n_k$. After an increase of 
the occupation of $n_{k=0}$ the system steadily relaxes to an 
approximately constant momentum distribution.

As before, in Fig.\ \ref{LatticeOff_PerfilK} we compare the results of the
time average of the momentum distribution function after the system has 
relaxed to a stationary distribution with the expectation 
values of this quantity in the grand-canonical (``thermal'' in the figure) 
and generalized (``constrained'' in the figure) Gibbs ensemble. 
Remarkably, when one starts from an insulating 
state, where correlations decay exponentially---i.e., there is no 
quasi-long-range order---the results obtained within a grand-canonical 
description are very similar to the ones obtained during the time evolution. 
As can be seen comparing Figs.\ \ref{LatticeOff_PerfilK}(a), 
\ref{LatticeOff_PerfilK}(b), and \ref{LatticeOff_PerfilK}(c), the deeper the 
initial state is in the insulating regime the better is the agreement between 
the grand-canonical and the average of the time evolving $n_k$'s after. 
Actually, only in Fig.\ \ref{LatticeOff_PerfilK}(a) 
can one clearly distinguish the differences in the occupation of the momentum 
states with very low values of $k$. On the other hand, the results obtained 
within the constrained theory are in all cases indistinguishable from the ones 
obtained during the time evolution.

\begin{figure}[h]
\begin{center}
\includegraphics[width=0.44\textwidth]
{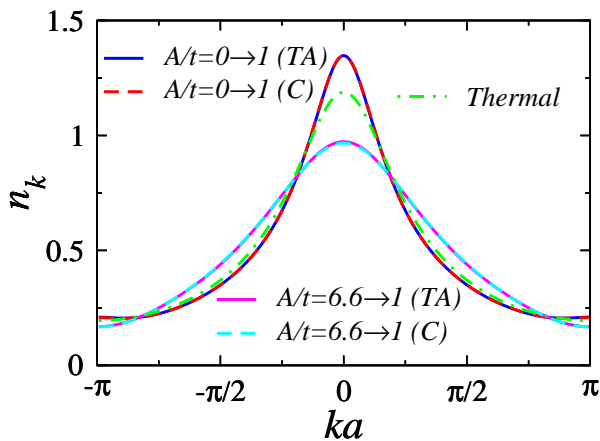}
\end{center} \vspace{-0.7cm}
\caption{(Color online) Comparison between the time average of the momentum 
distribution function of two systems starting from different 
initial conditions but having the same energy during the evolution.
(As in the previous figures, $n_k$ for the time average was measured in 
intervals $\Delta\tau=20t$ between times $\tau=20t$ and $\tau=5000t$.)
In both cases the evolution is done with a Hamiltonian in which 
$A=1.0t$, but the initial state is in one case a superfluid state
($A=0$), and in the other an insulating state with $A=6.6t$. 
The time average is compared with the results obtained 
in the usual thermal ensemble (which depends only on the final energy and
the number of particles---i.e., is the same in both cases) 
and the constrained theory explained in the text. 
These systems have 300 lattice sites, 150 particles, and $L=2$.
In the legend ``(TA)'' means time average in the 
out-of-equilibrium system and ``(C)'' the result of the constrained 
thermodynamics.}
\label{LatticeOnOff_PerfilK}
\end{figure}

A further proof of the ``memory'' that these integrable systems have on the 
initial conditions can be obtained comparing the final momentum distribution
function to which the system relaxes starting from two different initial 
states that have the same final energy. This comparison is done in 
Fig.\ \ref{LatticeOnOff_PerfilK} where we show results for two systems 
that start their evolution from a supperfluid (zero $A$) and insulating 
(nonzero $A$) states, evolving with a Hamiltonian in which the ground state 
is an insulator (nonzero $A$). While the usual grand-canonical description 
anticipates that they both should relax to the same momentum distribution, it 
can be seen in the figure that this is not the case. Only the generalized Gibbs 
distribution introduced before predicts two final momentum distribution 
functions and hence is able to describe the mean values of $n_k$ after 
relaxation.

To conclude this section we would like to make some remarks about the 
one-particle density matrix. While in the usual and generalized Gibbs 
distributions one can easily work in a basis where this quantity is 
real, this is not the case during the nonequilibrium dynamics of a 
quantum system. In the latter case, in all our calculations, 
the one-particle density matrix is a complex object 
($\rho_x(\tau)=|\rho_x(\tau)|\exp[-i\theta_x(\tau)]$) and 
nontrivial time-dependent phases [$\theta_x(\tau)$] enter into 
play. An example in which these phases play a fundamental role 
was mentioned in the Introduction. In 
Ref.\ \cite{rigol05_1} it was shown that during the free expansion 
of the hard-core boson gas on a lattice $n_k$ approaches the one of 
noninteracting fermions. This occurs due to the effects of 
$\theta_x(\tau)$ since $|\rho_x(\tau)|$ was found to decay with 
exactly the same power law than in equilibrium, which would mean 
that the system should have a large peak at $k=0$.

\begin{figure}[h]
\begin{center}
\includegraphics[width=0.44\textwidth]
{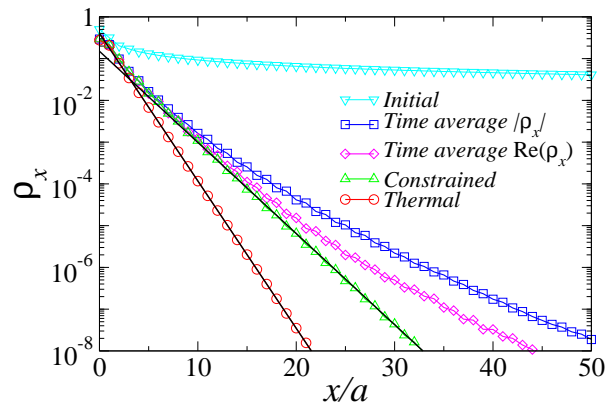}
\end{center} \vspace{-0.7cm}
\caption{(Color online) Time average of the modulus and real part of
the one-particle density matrix compared with the power-law decay 
present in the initial state and with the results obtained 
using the usual (thermal) and generalized (constrained) Gibbs 
distributions. We calculated $\rho_x$ from the center of the box, 
and for the time average we measured $\rho_x$ in time intervals 
$\Delta\tau=20t$ between $\tau=20t$ and $\tau=5000t$. These systems 
have 300 lattice sites, 150 particles, and $L=2$. The initial value of 
$A$ is $A=0$ and the final one $A=1.0t$. Straight lines following the
constrained and thermal results signal an exponential decay of $\rho_x$.}
\label{DensityMatrix_BOXDynamics}
\end{figure}

We find that after relaxation both the modulus and the real 
part of the one-particle density matrix in the time evolving system 
are very different from $\rho_x$ in the thermal and constrained ensemble, 
for large values of $x$. This can be seen in 
Fig.\ \ref{DensityMatrix_BOXDynamics} where 
we have plotted time average and thermodynamic results. [We did not 
include results for Im($\rho_x$) since it oscillates strongly between 
positive and negative values.] Figure \ref{DensityMatrix_BOXDynamics} shows 
that while in the usual and generalized Gibbs ensemble one-particle 
correlations decay exponentially at large distances (with different 
correlation lengths in each case), no such simple exponential decay 
can be seen either in the modulus or the real part of $\rho_x$ in the 
evolving system. In addition, the decay of the correlations in the latter 
case cannot be described by a power law like the one in the initial state. 
Their decay is actually faster. (We also show in the figure $\rho_x$ 
in the initial superfluid state.) The nontrivial effect of the 
phases $\theta_x(\tau)$ is evident in $n_k$, which is the diagonal part 
of the Fourier transform of the one-particle density matrix. While the 
mean value of $|\rho_x|$ in the out-of-equilibrium system decays more 
slowly than $\rho_x$ in the constrained thermodynamics, the $n_k$ 
obtained from both of them are identical [Fig.\ \ref{LatticeOn_PerfilK}(a)]. 
We then conclude that there is no simple comparison possible between the 
time average of one-particle correlations in the system out of equilibrium 
(after relaxation) and in its ``equivalent'' generalized Gibb's ensemble.
Hence, we do not consider that this quantity, which is not a physical 
observable, is an indicator of thermalization.

\section{HCB's in a superlattice plus 
a harmonic confining potential}

In this, and the following section, we generalize the results obtained 
in the previous sections to inhomogeneous systems. In particular, 
we study the effects of having a harmonic confining potential, relevant
to experimental systems, in addition to the superlattice potential. 
In this case the Hamiltonian reads
\begin{eqnarray}
\label{HamHCB_trap} H &=& -\sum_{i} \left( t_{i,i+1} b^\dagger_{i} b^{}_{i+1}
+ \mathrm{H.c.} \right) + A \sum_{i} \cos \left( {2 \pi i \over L} \right) 
n_i \nonumber \\ && + V_2 \sum_{i} x_i^2 \ n_{i },
\end{eqnarray}
where $V_{2}$ is the curvature of the harmonic trap. As in the previous 
sections in what follows we restrict the analysis to the case $L=2$.

From earlier studies of the bosonic Hubbard model in a trap it is 
known that in the presence of a confining potential superfluid and 
Mott-insulating phases coexist space separated
\cite{batrouni02,kollath04,wessel04}. On the other hand, 
in the HCB case it has been 
shown that the presence of the trap does not destroy the power-law 
decay of the one-particle correlations known from the homogeneous 
case \cite{rigol04_1}.

Previous studies have shown that the key parameter 
that controls the thermodynamic behavior of these confined systems 
is the characteristic density 
\begin{equation}
\tilde{\rho}=N_b/\zeta,
\end{equation}
where $\zeta=\left(V_{2}/t \right)^{-1/2}$ is a length 
scale set by the combination of the trap and the underlying 
lattice \cite{rigol04_1}. As $\tilde{\rho}\to 0$, one recovers 
the continuum limit. On the other hand, as $\tilde{\rho}$ increases
beyond a critical value insulating regions build up in the system.
In what follows we normalize distances in the trap by $\zeta$ and
calculate the momentum distribution function as
\begin{equation}
n_k=\frac{a}{\zeta}\sum_{jl} e^{-ik(j-l)}\rho_{jl}.
\end{equation}

\begin{figure}[h]
\begin{center}
\includegraphics[width=0.48\textwidth]
{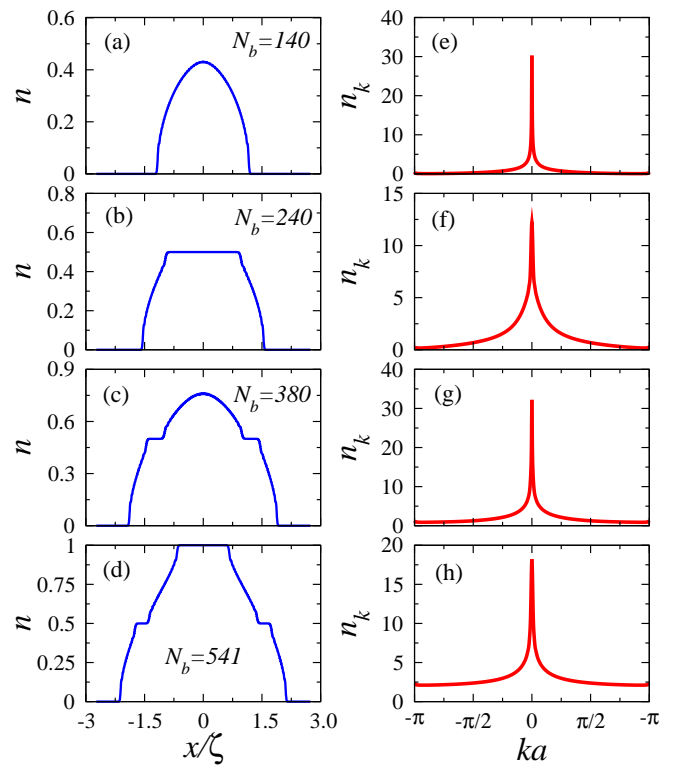}
\end{center} \vspace{-0.7cm}
\caption{(Color online) Density (averaged per unit cell) (a)--(d)
and normalized momentum distribution function (e)--(h) 
for trapped systems with 1000 lattice sites, 
$V_2a^2=3\times 10^{-5}t$, $L=2$, and $A=0.5t$. 
In (a)--(d), the regions with constant density are local insulators.}
\label{Perfiles_Trap}
\end{figure}

In Figs.\ \ref{Perfiles_Trap}(a)--\ref{Perfiles_Trap}(d) we show 
the density profiles (averaged per unit cell) for harmonically 
trapped systems with $A=0.5t$, at different fillings. These plots 
show that like in the Bose-Hubbard case superfluid and insulating 
(constant density $n=0.5$ and 1) phases coexist 
space separated in the trap. The width of the insulating phases 
with $n=0.5$ is determined by the gap, which for the cases depicted in 
Fig.\ \ref{Perfiles_Trap} is $\Delta=t$. The effect of the curvature
of the harmonic potential is already considered by normalizing 
the distances by $\zeta$.

\begin{figure}[h]
\begin{center}
\includegraphics[width=0.44\textwidth]
{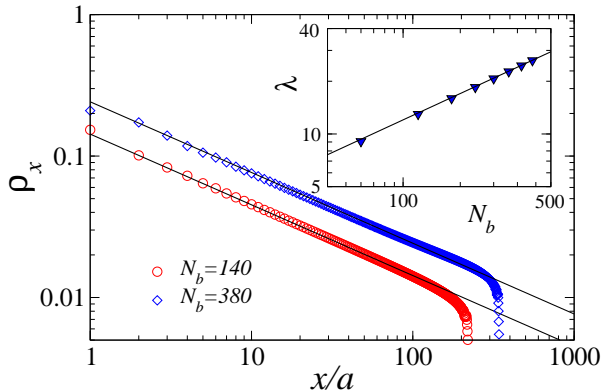}
\end{center} \vspace{-0.7cm}
\caption{(Color online) Decay of one-particle correlations $\rho_x$
($x=|x_i-x_j|$) (averaged per unit cell as described in Sec.\ II) 
in the superfluid domains. We have chosen $x_i$ to be the center of the trap
[see the corresponding density profiles in 
Figs.\ \ref{Perfiles_Trap}(a) and \ref{Perfiles_Trap}(c)]. 
Straight lines signal a power-law decay $\rho_x\sim 1/\sqrt{x}$, 
which disappears when $n_j\rightarrow 0,\,0.5,$ or 1. 
In the inset we show the scaling of the lowest natural orbital occupation
with an increasing number of particles keeping constant the characteristic
density of the system, which in this case is $\tilde{\rho}=0.66$. 
The straight line shows that $\lambda_0\sim \sqrt{N_b}$.}
\label{DensityMatrix_Trap}
\end{figure}

The formation of insulating domains in the trap strongly
supresses the peaks observed in the momentum distribution at
$k=0$. This can be seen in 
Figs.\ \ref{Perfiles_Trap}(e)--\ref{Perfiles_Trap}(h), 
where we have plotted the momentum profiles corresponding
to the densities in 
Figs.\ \ref{Perfiles_Trap}(a)--\ref{Perfiles_Trap}(d). 
However, these peaks are not destroyed and can be very sharp. 
As can be seen in Fig.\ \ref{DensityMatrix_Trap}, the 
reason is that quasi-long-range correlations are still present
in the superfluid regions.  We find that their power-law decay 
is the same observed in the absence of the superlattice 
\cite{rigol04_1} and in the periodic case discussed in 
Sec.\ II---i.e., $\rho_x\sim 1/\sqrt{x}$, with $x=|x_i-x_j|$. 
Here of course translational invariance is broken by the trap; 
however, the above power law can be seen independently of the 
points $x_i$ and $x_j$ chosen within the superfluid part. 
On the other hand, as expected, in the insulating phases, 
where the density is constant, the decay of $\rho_x$ is 
exponential, exactly like the one in 
Fig.\ \ref{DensityMatrix_Insulator}. The particles there, being 
localized, are the ones that mainly contribute to the high momentum 
tails of $n_k$.

More information about the physics of the trapped system can be 
obtained studying the natural orbitals ($\phi^\eta$), which can be 
considered to be like effective single-particle states when the 
system consists of interacting particles. They are defined as 
the eigenfunctions of the one-particle density matrix
$\rho_{ij}$ \cite{penrose56},
\begin{equation}
\label{NatOrb}
\sum^N_{j=1} \rho_{ij}\phi^\eta_j=
\lambda_{\eta}\phi^\eta_i.
\end{equation}
$\lambda_{\eta}$ denotes the occupation of each orbital. In higher 
dimensions, when only the lowest natural orbital (the highest occupied one) 
scales $\sim N_b$, it can be regarded as the BEC order parameter---i.e., 
the condensate \cite{leggett01}. This scaling of the lowest natural orbital
in higher dimensions is related to the presence of long-range off-diagonal 
order \cite{yang62}. In the one-dimensional case we are studying
here there is no long-range order, only quasi-long-range order 
characterized by a $1/\sqrt{x}$ decay of the correlations 
(Fig.\ \ref{DensityMatrix_Trap}), which as shown in the inset of 
Fig.\ \ref{DensityMatrix_Trap} produces a power-law
($\sqrt{N_b}$) scaling of the lowest natural orbital occupation.
(These scaling laws have also been observed in harmonically trapped 
continuous systems \cite{papenbrock03,forrester03,gangardt04}.) 
Hence, we refer to the lowest natural orbital as a 
quasicondensate since $\lambda_0\rightarrow\infty$ when 
$N_b\rightarrow\infty$, but $\lambda_0/N_b\rightarrow 0$.

\begin{figure}[h]
\begin{center}
\includegraphics[width=0.48\textwidth]
{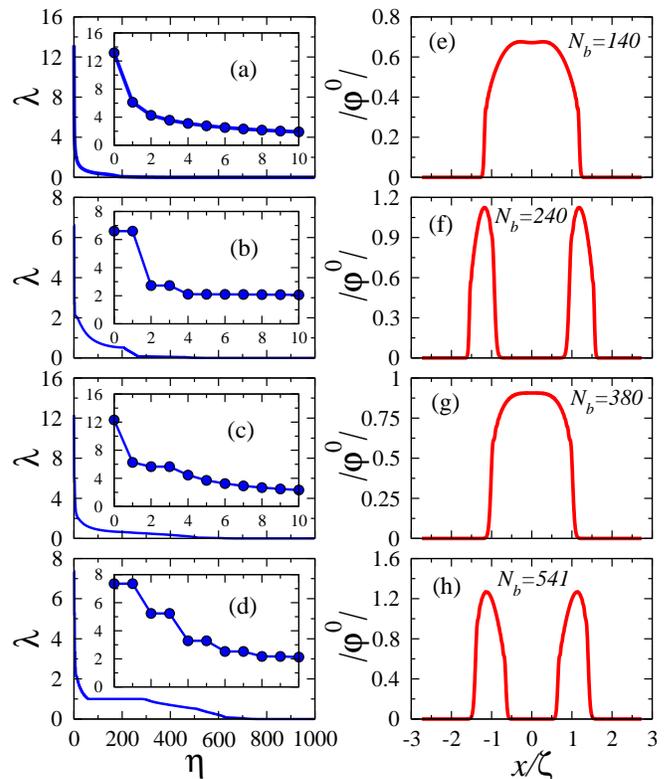}
\end{center} \vspace{-0.7cm}
\caption{(Color online) Natural orbital occupations (a)--(d) 
and normalized wave function of the lower natural orbital 
$\varphi^0=\left( N_b \zeta/ a\right)^{1/4} \phi^0$ 
(averaged per unit cell) (e)--(h), for trapped systems with 1000 
lattice sites, $V_2a^2=3\times 10^{-5}t$, $L=2$, and $A=0.5t$.
In the insets of (a)--(d) we show the occupation of the lowest 11
natural orbitals, where degeneracy can be seen in the presence
of the insulating domains. Comparing (e)--(h) with 
Figs.\ \ref{Perfiles_Trap}(a)--\ref{Perfiles_Trap}(d) 
one can see that the lowest natural orbitals only have a 
finite weight in the superfluid domains.}
\label{NatOrb_Trap}
\end{figure}

In Figs.\ \ref{NatOrb_Trap}(a)--\ref{NatOrb_Trap}(d) we show the 
occupation of the natural orbitals (ordered from the highest occupied 
one to the lowest occupied one) as a function of the orbital number 
$\eta$. The formation of the insulating domains not only reduces 
the occupation of the lowest natural orbitals, but also makes them 
degenerate (insets). The normalized wave function of the 
lowest natural orbitals \cite{rigol04_1}
\begin{equation}
\varphi^0=\left( N_b \zeta/ a\right)^{1/4} \phi^0,
\end{equation} 
are plotted in Figs.\ \ref{NatOrb_Trap}(e)--\ref{NatOrb_Trap}(h). 
Comparing these wave functions with the density profiles in 
Figs.\ \ref{Perfiles_Trap}(a)--\ref{Perfiles_Trap}(d) one can see 
that these orbitals are localized in the superfluid regions and have 
zero weight in the insulating domains. Degeneracy then appears because 
two identical quasicondensates develop to the sides of the central 
insulating core in the cases depicted in 
Figs.\ \ref{Perfiles_Trap}(b) and \ref{Perfiles_Trap}(d). According 
to the number of particles in the system the lowest natural orbitals 
can be located outside or inside the insulating domains with average 
density $n=0.5$. 

\begin{figure}[h]
\begin{center}
\includegraphics[width=0.44\textwidth]
{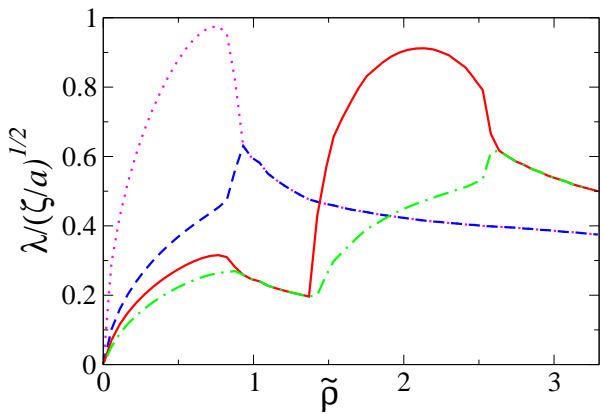}
\end{center} \vspace{-0.7cm}
\caption{(Color online) Occupations of the lowest 4 natural orbitals
as a function of the characteristic density. These systems have 1000 
lattice sites, $V_2a^2=3\times 10^{-5}t$, $L=2$, and $A=0.5t$.
Degeneracy sets in when insulating domains (with density $n=0.5$ first and
$n=1.0$ second) appear in the middle of the trap.}
\label{lambdavsNb}
\end{figure}

The behavior of the lowest natural orbital occupations with an increasing 
number of particles in the trap can be seen in Fig.\ \ref{lambdavsNb}.
It perfectly reflects the formation and destruction of insulating domains
in the system. The first degeneracy for the four lowest natural orbitals 
(Fig.\ \ref{lambdavsNb}) appear when the insulating phase with mean 
density $n=0.5$ sets in the 
middle of the trap [Fig.\ \ref{Perfiles_Trap}(b)]. This degeneracy 
disappears for the natural orbitals 3 and 4 when a superfluid 
domain, like the one in Fig.\ \ref{Perfiles_Trap}(c), develops in the 
center of the system. The central superfluid phase grows with increasing 
number of particles, and the quasicondensate there becomes the 
highest populated. Finally, degeneracy sets up once again when the full 
filled insulator ($n=1$) appears in the center of the trap 
[like in Fig.\ \ref{Perfiles_Trap}(d)].

\section{Dynamics and relaxation in a harmonic trap}

In this section we study the dynamics in the presence of a harmonic trap.
Since at very low densities in the trap, when the average interparticle 
distance is much larger than the lattice spacing, the lattice system 
and the continuum are equivalent, we start analyzing this case. 

\begin{figure}[h]
\begin{center}
\includegraphics[width=0.46\textwidth]
{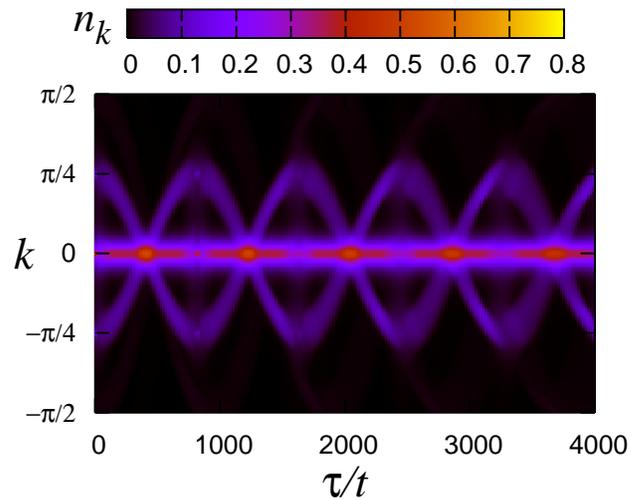}
\end{center} \vspace{-0.7cm}
\caption{(Color online) Evolution of the momentum distribution
function after a superlattice potential with $L=8$ and $A=2.0t$ 
is applied for a short period of time ($\Delta\tau=0.5t$) 
to the ground state of trapped system with 10 HCB's, 
$V_2a^2=4\times 10^{-6}t$, and 2000 lattice sites. Undamped 
oscillations can be seen in $n_k$ during our simulation time.}
\label{continuous3D}
\end{figure}

We perform a numerical experiment similar to that recently done at 
Penn State \cite{kinoshita06}. There, in order to generate a highly 
exited state in a trap, an optical lattice potential was applied 
during a short period of time to an array of 1D Bose gases. 
This produced a momentum distribution with extra peaks away from $k=0$ 
(the only one present in the absence of a lattice). After turning off the 
lattice potential the dynamics of the system was studied using 
time-of-flight measurements. It was found that after some periods of 
oscillation the system relaxed to a momentum distribution that was not 
the one in thermal equilibrium. Relaxation to an equilibrium distribution
was attributed mainly to the effects of the anharmonicity of the 
confining potential. In Ref.\ \cite{rigol06_1} and in this 
paper, we have shown why in the integrable limit of infinitely strong 
interactions the system relaxes to a distribution that is not the one 
in thermal equilibrium. Surprisingly, in the experiments \cite{kinoshita06} 
it was found that the absence of thermalization extends towards the 
nonintegrable region of finite interactions. Even though these 1D systems
can be very well described by the Lieb-Liniger model \cite{kinoshita04},
which is integrable in periodic homogeneous systems, the presence of
the trapping potential in the experiment breaks integrability but did 
not allow the system to thermalize.

To recreate the experiment in Ref.\ \cite{kinoshita06} (performed in 
continuous space) in a perfect harmonic potential, we consider a very 
dilute system in our lattice model. We study the dynamics of 10 HCB's 
in 2000 lattice sites; i.e., the interparticle distance is much larger 
than the lattice spacing so that the effects of the underlying lattice 
are negligible. We then turn on, for a short period of time, an additional 
periodic potential with $L=8$. The dynamics of the momentum distribution 
function after turning off this additional lattice is shown in 
Fig.\ \ref{continuous3D}. This figure shows that no relaxation towards 
an equilibrium distribution can be seen in the perfectly harmonic 
case. The additional peaks in $n_k$ oscillate back and forth 
almost without damping. This occurs even when many eigenmodes
of the system have been excited by turning on and off the lattice 
potential. Our results confirm the conclusion in Ref.\ \cite{kinoshita06} 
that relaxation in the hard-core regime may be mainly related to the 
anharmonicity of the confining potential.

While the low-energy region (low occupation) of the harmonic 
trap spectrum in the presence of the lattice is like the one 
in the continuum, this is not true anymore in the high-energy 
region (high occupation); i.e., the lattice model and 
the continuum one start to differ \cite{rigol04_1,cazalilla04}. 
The spectrum in the lattice departs from the linear dispersion relation 
of the harmonic potential, and degeneracies set in when insulating 
domains develop in the ground state \cite{rigol04_3}. Hence, for 
large fillings in the trap we may expect to see relaxation towards 
an equilibrium distribution.

\begin{figure}[h]
\begin{center}
\includegraphics[width=0.44\textwidth]
{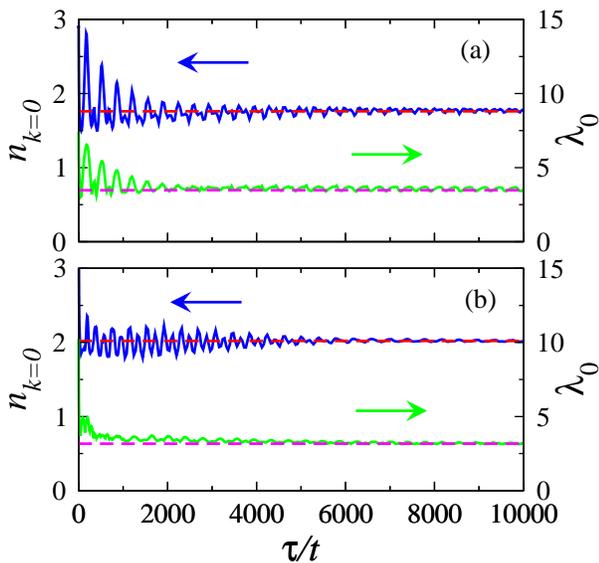}
\end{center} \vspace{-0.7cm}
\caption{(Color online) Evolution of the occupation of the 
zero-momentum state (top plots) and the occupation of the
lowest natural orbital (bottom plots) after a 
superlattice potential is turned off in trapped systems
with 900 lattice sites and $V_2a^2=3\times 10^{-5}t$. 
The evolution starts from the ground state of a system with 
$L=2$ and $A=0.5t$. The number of particles is $N_b=200$ (a)
and $N_b=299$ (b). The corresponding initial density profiles 
can be seen in Figs.\ \ref{TrapLattOff_Perfiles}(a) and  
\ref{TrapLattOff_Perfiles}(d). The dashed lines are the results 
obtained with the generalized Gibbs distribution explained 
in the text.}
\label{TrapLattOff_nk0no0vstau}
\end{figure}

\begin{figure}[h]
\begin{center}
\includegraphics[width=0.485\textwidth]
{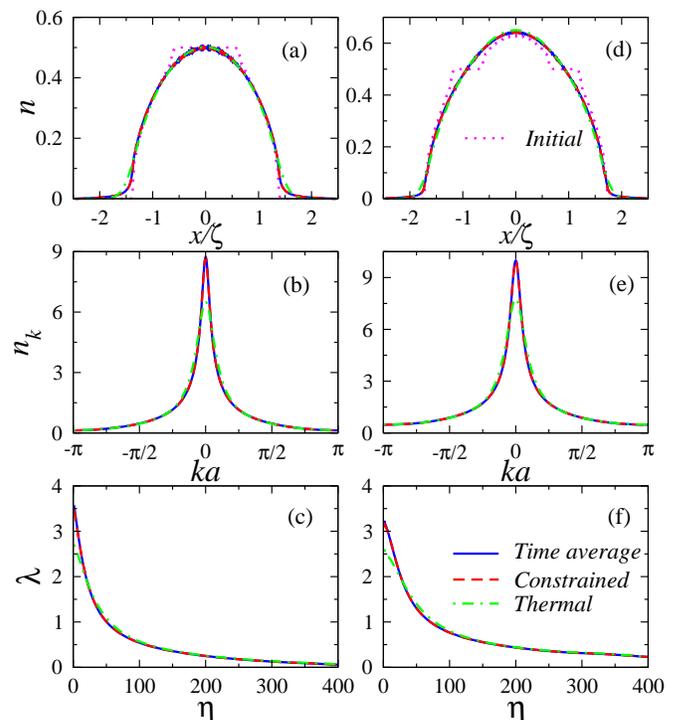}
\end{center} \vspace{-0.7cm}
\caption{(Color online) Time average of the 
density profiles (a),(d), momentum distribution function 
(b),(e), and the occupation of the lowest 400 natural orbitals. 
The average is performed between $\tau=5000t$ and $\tau=10000t$ 
(see Fig.\ \ref{TrapLattOff_nk0no0vstau}) with measurements
done in time intervals $\Delta\tau=40t$. The evolution is 
performed in trapped systems with 900 lattice sites, 
$V_2a^2=3\times 10^{-5}t$, starting from the ground state 
in the presence of a superlattice with $L=2$ and $A=0.5t$, 
after turning off the superlattice. The results of the 
time average are compared with the ones obtained in the usual 
thermal ensemble and the constrained theory explained in the 
text. The number of particles is $N_b=200$ (a)--(c)
and $N_b=299$ (d)--(f). In (a) and (d) we included the averaged
density per unit cell in the initial state. Flat regions 
correspond to insulating domains.}
\label{TrapLattOff_Perfiles}
\end{figure}

We study the dynamics of systems in which initially, 
in the presence of a superlattice potential, there is a coexistence 
of superfluid and insulating domains. We then turn off the superlattice 
potential and let the system evolve in the presence of the trap.
In Fig.\ \ref{TrapLattOff_nk0no0vstau} we show the evolution of
the occupations of the zero-momentum state ($n_{k=0}$) and the lowest 
natural orbital ($\lambda_0$) when (a) the initial state has a half-filled
insulator in the center of the trap [Fig.\ \ref{TrapLattOff_Perfiles}(a)]
and (b) two insulating shoulders surround a central superfluid region
[Fig.\ \ref{TrapLattOff_Perfiles}(d)]. Figure \ref{TrapLattOff_nk0no0vstau}
shows that for high fillings there is a strong damping of the oscillations 
of $n_{k=0}$ and $\lambda_0$, in contrast to the harmonically trapped case 
without the lattice. Hence, in experiments relaxation to an equilibrium 
state is to be expected when a lattice is present along the 1D tubes.  

As seen in Fig.\ \ref{TrapLattOff_nk0no0vstau} after $\tau=5000t$
the oscillations of $n_{k=0}$ and $\lambda_0$ are very small and occur 
around an approximately constant value. In Fig.\ \ref{TrapLattOff_Perfiles},
we compare the time average (between $\tau=5000$ and $10000t$) 
of the density profiles, the momentum distribution function, 
and the natural orbitals, with the predictions of the usual 
grand-canonical ensemble and the generalized Gibbs distribution
introduced in Ref.\ \cite{rigol06_1}. For all quantities the results
of the time averages and the generalized Gibbs distribution are on
top of each other. On the other hand, the differences between time 
averages and the thermal ensemble are apparent in all cases and 
particularly large in the low-momentum region of $n_k$ and in the 
occupation of the lowest natural orbitals. Only in the density 
profiles are the differences smaller and mainly visible in the 
regions where the density approaches zero.

Since the density profiles of HCB's and noninteracting fermions are 
identical and their time average coincides with the results of the
generalized Gibbs distribution, one realizes that noninteracting 
systems are the simplest case to which the constrained thermal 
equilibrium theory \cite{rigol06_1} explained in Sec.\ III can be 
applied.

\section{Conclusions}

We have studied ground-state properties and the nonequilibrium 
dynamics of hard-core bosons on one-dimensional lattices in the 
presence of an additional periodic potential (superlattice)
and a harmonic trap.

In the periodic case the superlattice potential opens gaps at the
borders of the reduced Brillouin zones generating insulating phases
with fractional fillings. In these insulating phases one-particle
correlations decay exponentially $[\rho_x\sim \exp(-x/\xi)]$, and 
we have studied how the correlation length $\xi$ behaves as a 
function of the strength $A$ of the superlattice potential. 
On the other hand, we find that in the gapless superfluid phases 
one-particle correlations exhibit quasi-long-range order. They 
decay with the same power law shown in Ref.\ \cite{rigol04_1} 
to be universal in the absence of the superlattice potential. 

In the presence of an additional confining potential, which we 
have taken to be harmonic for a closer connection 
to the experiments, superfluid and insulating domains coexist 
phase separated. We have shown that in the superfluid domains, 
where the density changes spatially, one-particle correlations 
decay with the same power law than in the homogeneous periodic 
case. This decay produces a square-root scaling of the 
occupation of the lowest natural orbital with the number of 
particles in the trap and independently of the presence or not 
of insulating domains.

We have also studied the nonequilibrium dynamics of these systems
after a quench of the superlattice potential $A$. In particular, 
we have considered a half-filled box in the presence of a superlattice 
with period 2, since for $A=0$ the system is superfluid while 
for $A\neq 0$ it is insulating. After a sudden switch-on of $A$ we 
have shown that the initial momentum distribution ($n_k$) of the 
superfluid phase collapses and revives with a period determined by $A$, 
like in the experiment in Ref.\ \cite{greiner02} where the period was 
determined by the on-site repulsion $U$. After several oscillations, 
the number depending on $A$ and the hopping parameter $t$, we have 
also seen that the system relaxes to an equilibrium distribution with 
very small fluctuations in $n_k$. The time average of this physical 
observable was then shown to be very well described by a generalized
Gibbs distribution introduced in Ref.\ \cite{rigol06_1}. On the other
hand, after a sudden switch-off of $A$ we have found that not only the
generalized Gibbs ensemble but also the usual thermal ensemble describes 
very well the time average result when the system is initially deep 
in the insulating regime ($A>t$). This is indeed a surprising result since
we are seeing thermalization in an integrable system. But of course it is
only seen in the very particular case in which the quench of the 
interaction drives the system from an insulating state to a superfluid 
one. In the latter the temperature plays a very similar role than the gap 
present in the initial insulator. In transitions between different 
superfluid states \cite{rigol06_1} no such thermalization is seen.

Finally, we have considered the nonequilibrium dynamics for the 
harmonically confined case. We have shown that in this more experimentally 
relevant system the time average of observables, like density and momentum 
distribution, can also be very well described by the generalized Gibbs 
distribution. Damping occurs in this case, even in a perfect harmonic trap, 
because of the presence of the lattice. At very low densities, equivalent to 
the continuum case, we have shown that oscillations of $n_k$ occur almost 
without damping.

\begin{acknowledgments}
We would like to thank the Max Planck Institute for the Physics of Complex 
Systems in Dresden for the hospitality during the ``Workshop on Non-equilibrium 
Dynamics in Interacting System'' where this work was initiated. We acknowledge
stimulating conversations during the workshop with M. A. Cazalilla, C. Kollath, 
A. Kolovsky, S. Manmana, D. Weiss, and X. Zotos. We are also grateful 
to R.~T. Scalettar and R.~R.~P. Singh for helpful discussions. This work was 
supported by National Science Foundation grants Nos. NSF-DMR-0240918, 
NSF-DMR-0312261, NSF-PHY-0301052, a grant from Office of Naval Research 
No. N00014-03-1-0427, and SFB/TR 21 and HLR-Stuttgart in Germany.
\end{acknowledgments}


\begin{thebibliography}{99}

\bibitem{dresden06} http://www.mpipks-dresden.mpg.de/$\sim$neqdis06/

\bibitem{thyw99} J.~H. Thywissen, R.~M. Westervelt, and M. Prentiss,
Phys. Rev. Lett. {\bf 83}, 3762 (1999).

\bibitem{muller99} D. M\"uller, D.~Z. Anderson, R.~J. Grow, P.~D.~D.
Schwindt, and E.~A. Cornell, Phys. Rev. Lett. {\bf 83}, 5194 (1999).

\bibitem{dekker00} N.~H. Dekker, C.~S. Lee, V. Lorent, J.~H. Thywissen,
S.~P. Smith, M. Drndic, R.~M. Westervelt, and M. Prentiss, Phys.
Rev. Lett. {\bf 84}, 1124 (2000).

\bibitem{key00} M. Key, I.~G. Hughes, W. Rooijakkers, B.~E. Sauer,
E.~A. Hinds, D.~J. Richardson, and P.~G. Kazansky, Phys. Rev. Lett. {\bf 84},
1371 (2000).

\bibitem{bongs01} K. Bongs, S. Burger, S. Dettmer, D. Hellweg, J. Arlt,
W. Ertmer, and K. Sengstock, Phys. Rev. A {\bf 63}, 031602(R) (2001).

\bibitem{schreck01} F. Schreck, L. Khaykovich, K.~L. Corwin, G. Ferrari,
T. Bourdel, J. Cubizolles, and C. Salomon, Phys. Rev. Lett. {\bf 87},
080403 (2001).

\bibitem{gorlitz01} A. G\"orlitz, J.~M. Vogels, A.~E. Leanhardt, C. Raman,
T.~L. Gustavson, J.~R. Abo-Shaeer, A.~P. Chikkatur, S. Gupta, S. Inouye, 
T. Rosenband, and W. Ketterle, Phys. Rev. Lett. {\bf 87}, 130402 (2001).

\bibitem{strecker02} K.~E. Strecker, G.~B. Partridge,
A.~G. Truscott, and R.~G. Hulet, Nature (London), {\bf 417}, 150 (2002).

\bibitem{greiner01} M. Greiner, I. Bloch, O. Mandel, T.~W. H\"ansch, and
T. Esslinger, Phys. Rev. Lett. {\bf 87}, 160405 (2001).

\bibitem{moritz03} H. Moritz, T. St\"oferle, M. K\"ohl, and T. Esslinger, 
Phys. Rev. Lett. {\bf 91}, 250402 (2003). 

\bibitem{stoferle03} T. St\"oferle, H. Moritz, C. Schori, M. K\"ohl, and
T. Esslinger, Phys. Rev. Lett. {\bf 92}, 130403 (2004).

\bibitem{tolra04} B.~L. Tolra, K.~M. O'Hara, J.~H. Huckans, W.~D. Phillips, 
S.~L. Rolston, and J.~V. Porto, Phys. Rev. Lett. {\bf 92}, 190401 (2004).

\bibitem{paredes04} B. Paredes, A. Widera, V. Murg, 
O. Mandel, S. F\"olling, I. Cirac, G.~V. Shlyapnikov, 
T.~W. H\"ansch, and I. Bloch, Nature (London) {\bf 429}, 277 (2004).

\bibitem{kinoshita04} T. Kinoshita, T. Wenger, and D.~S. Weiss, 
Science {\bf 305}, 1125 (2004).

\bibitem{fertig05} C.~D. Fertig, K.~M. O'Hara, J.~H. Huckans, 
S.~L. Rolston, W.~D. Phillips, and J.~V. Porto,
Phys. Rev. Lett. {\bf 94}, 120403 (2005). 

\bibitem{kinoshita06}
T. Kinoshita, T. Wenger, and D.~S. Weiss, 
Nature (London) {\bf 440}, 900 (2006).

\bibitem{rigol06_1}
M. Rigol, V. Dunjko, V. Yurovsky, and M. Olshanii, 
e-print cond-mat/0604476.

\bibitem{igloi00} F. Igl\'oi and H. Rieger,
Phys. Rev. Lett. {\bf 85}, 3233 (2000).

\bibitem{sengupta04} K. Sengupta, S. Powell, and S. Sachdev,
Phys. Rev. A {\bf 69}, 053616 (2004).

\bibitem{calabrese06} P. Calabrese and J. Cardy,
Phys. Rev. Lett. {\bf 96}, 136801 (2006).

\bibitem{cherng06} R. W. Cherng and L. S. Levitov,
Phys. Rev. A {\bf 73}, 043614 (2006).

\bibitem{cazalilla06} M.~A. Cazalilla, 
Phys. Rev. Lett. {\bf 97}, 156403 (2006).

\bibitem{olshanii98} M. Olshanii, 
Phys. Rev. Lett. {\bf 81}, 938 (1998).

\bibitem{petrov00} D.~S. Petrov, G.~V. Shlyapnikov, 
and J.~T.~M. Walraven, Phys. Rev. Lett. {\bf 85}, 3745 (2000).

\bibitem{dunjko01} V. Dunjko, V. Lorent, and M. Olshanii, 
Phys. Rev. Lett. {\bf 86}, 5413 (2001).

\bibitem{girardeau60} M. Girardeau, J. Math. Phys. {\bf 1}, 516 (1960).

\bibitem{girardeau00} M.~D. Girardeau and E.~M. Wright, 
Phys. Rev. Lett. {\bf 84}, 5691 (2000).

\bibitem{girardeau00a} M.~D. Girardeau and E.~M. Wright, 
Phys. Rev. Lett. {\bf 84}, 5239 (2000). 

\bibitem{girardeau02} K.~K. Das, M.~D. Girardeau, and E.~M. Wright, 
Phys. Rev. Lett. {\bf 89}, 170404 (2002).

\bibitem{minguzzi05} A. Minguzzi and D.~M. Gangardt,
Phys. Rev. Lett. {\bf 94}, 240404 (2005).

\bibitem{campo05} A. del Campo and J.~G. Muga, 
Europhys. Lett. {\bf 74}, 965 (2006).

\bibitem{lieb61} E. Lieb, T. Shultz, and D. Mattis, 
Ann. Phys. (N.Y.) {\bf 16}, 407 (1961).

\bibitem{mccoy68} B.~M. McCoy, Phys. Rev. {\bf 173}, 531 (1968).

\bibitem{vaidya78} H.~G. Vaidya and C.~A. Tracy, 
Phys. Lett. {\bf 68}A,  378 (1978).

\bibitem{mccoy83} B.~M. McCoy, J.~H. Perk, and R.~E. Schrock, 
Nucl. Phys. B {\bf 220}, 35 (1983); {\bf 220}, 269 (1983).

\bibitem{gangardt06} D.~M. Gangardt and G.~V. Shlyapnikov, 
New J. Phys. {\bf 8} 167 (2006).

\bibitem{rigol04_1} M. Rigol and A. Muramatsu, 
Phys. Rev. A {\bf 70}, 031603(R) (2004); {\bf 72}, 013604 (2005).

\bibitem{rigol04_2} M. Rigol and A. Muramatsu,
Phys. Rev. Lett. {\bf 93}, 230404 (2004).

\bibitem{rigol05_1} M. Rigol and A. Muramatsu,
Phys. Rev. Lett. {\bf 94}, 240403 (2005).

\bibitem{rigol05_2} M. Rigol and A. Muramatsu,
Mod. Phys. Lett. B {\bf 19}, 861 (2005).

\bibitem{peil03}
S. Peil, J.~V. Porto, B. L. Tolra, J.~M. Obrecht, B.~E. King, 
M. Subbotin, S.~L. Rolston, and W.~D. Phillips,
Phys. Rev. A {\bf 67}, 051603(R) (2003).

\bibitem{strabley06} J. Sebby-Strabley, M. Anderlini, P.~S. Jessen, 
and J.~V. Porto, Phys. Rev. A {\bf 73}, 033605 (2006).

\bibitem{buonsante04} P. Buonsante and A. Vezzani,
Phys. Rev. A {\bf 70}, 033608 (2004); 
P. Buonsante, V. Penna, and A. Vezzani, 
{\it ibid.} {\bf 70}, 061603(R) (2004);
P. Buonsante and A. Vezzani,
{\it ibid.} {\bf 72}, 013614 (2005).

\bibitem{rousseau06} V.~G. Rousseau, D.~P. Arovas, M. Rigol, 
F. H\'ebert, G.~G. Batrouni, and R.~T. Scalettar,
Phys. Rev. B {\bf 73}, 174516 (2006).

\bibitem{rey06} A.~M. Rey, I.~I. Satija, and C.~W. Clark,
New J. Phys. {\bf 8}, 155 (2006).

\bibitem{aizenman04} M. Aizenman, E.~H. Lieb, R. Seiringer,
J.~P. Solovej, and J. Yngvason, Phys. Rev. A {\bf 70}, 023612 (2004).

\bibitem{jordan28} P. Jordan and E. Wigner, 
Z. Phys. {\bf 47}, 631 (1928).

\bibitem{rigol05_3} M. Rigol, Phys. Rev. A {\bf 72}, 063607 (2005).

\bibitem{greiner02} M. Greiner, O. Mandel, T.~W. H\"ansch, and 
I. Bloch, Nature (London) {\bf 419}, 51 (2002).

\bibitem{batrouni02}
G.~G. Batrouni, V. Rousseau, R.~T. Scalettar, M. Rigol, A. Muramatsu, 
P.~J.~H. Denteneer, and M. Troyer, Phys. Rev. Lett. {\bf 89}, 117203 (2002).

\bibitem{kollath04}
C. Kollath, U. Schollw\"ock, J. von Delft, and W. Zwerger, 
Phys. Rev. A {\bf 69}, 031601(R) (2004).

\bibitem{wessel04}
S. Wessel, F. Alet, M. Troyer, and G.~G. Batrouni, 
Phys. Rev. A {\bf 70}, 053615 (2004).

\bibitem{penrose56} O. Penrose and L. Onsager, 
Phys. Rev. {\bf 104} 576, (1956).

\bibitem{leggett01}
A.~J. Leggett, Rev. Mod. Phys. {\bf 73}, 307 (2001).

\bibitem{yang62} C.~N. Yang, Rev. Mod. Phys. {\bf 34}, 694 (1962).

\bibitem{papenbrock03} T. Papenbrock, Phys. Rev. A {\bf 67}, 041601(R) (2003).

\bibitem{forrester03} P. J. Forrester, N. E. Frankel, T. M. Garoni,
and N. S. Witte, Phys. Rev. A {\bf 67}, 043607 (2003).

\bibitem{gangardt04} D. M. Gangardt, 
J. Phys. A {\bf 37}, 9335 (2004).

\bibitem{cazalilla04} M. A. Cazalilla,
Phys. Rev. A {\bf 70}, 041604(R) (2004).

\bibitem{rigol04_3}  M. Rigol and A. Muramatsu, 
Phys. Rev. A {\bf 70}, 043627 (2004).


\end{thebibliography}
\end{document}